\documentclass[pra,article]{revtex4}

\newcommand{\re}{\operatorname{Re}\,}
\newcommand{\im}{\operatorname{Im}\,}

\newcommand{\rme}{\operatorname{\rm e}}
\newcommand{\rmd}{\operatorname{\rm d}}
\newcommand{\rmi}{\operatorname{\rm i}}

\newcommand{\sgn}{\operatorname{sgn}\,}

\usepackage{amsfonts}
\usepackage{amssymb}
\usepackage{amsmath}
\usepackage{bbm}
\usepackage[]{graphicx}

\begin{document}

\title{\Large Quantum transmission for embedded, locally periodic potentials $-$ amplitude-phase approaches}

\author{Karl-Erik Thylwe $\vspace {0.3cm}$}

\affiliation { Department of Mechanics, KTH- Royal Institute of Technology, S-10044 Stockholm, Sweden }
\begin{abstract}
Quantum particle transmission through locally periodic potentials surrounded by symmetric exterior potentials is analyzed.  Closed-form conditions for locating energy peaks of total transmission are derived. Floquet/Bloch energy band types are defined and found to affect the number of peaks in transmission bands. Modifications of band types and a band fusion phenomenon are discussed. The theoretical approaches suggested consist of several ways to express a Schr\"{o}dinger wave as an amplitude function multiplying a harmonic function of a phase function.  In particular, an approach containing a Floquet/Bloch periodic amplitude function and a corresponding phase function is described for the locally periodic region, allowing a way to specify quantum numbers for energies of total transmission.
\end{abstract}
\maketitle

\section{Introduction}
One-dimensional quantum scattering caused by locally periodic potentials introduce useful theoretical notions like Floquet/Bloch energy bands and gaps, which are helpful for understanding more complicated physical systems. Chemical selections of gas components \cite{Mandrˆ14,Pereyra09} is one application. For example mass selections, where total transmission is monitored for a certain mass and minimal transmission for a slightly different mass.  Manipulation of electronic properties of material structures such as graphene \cite{ReviewsG}-\cite{Cury87} is another of many  applications. An introductory review is given by Griffiths and Steinke  (2001) \cite{Griffiths01}. The  authors cover examples  related to several mechanical systems: transverse waves on weighted strings, longitudinal waves on weighted rods, acoustic waves in corrugated tubes, and water waves crossing a sequence of sandbars. The authors also discuss electromagnetic waves in transmission lines and photonic crystals, as well as relativistic quantum scattering described by the Dirac equation in one space dimension.
The so-called 'transfer matrix' method \cite{Griffiths01,Yu90,Sprung93} is applied in most published numerical studies. A limitation is possibly the lack of realistic interactions studied. Potentials are usually constructed by delta functions and/or square well/barrier functions of the space coordinate \cite{Griffiths01,Dharani16,Yu17}.  Novel aspects of {\it non-uniform} lattice potentials constructed by various rectangular potential pieces are discussed by Das (2015) \cite{Das15}. In the present study the locally periodic part of the potentials are {\it uniform} and smooth. Non-smooth potentials cannot easily be generalized to second-order Dirac equations for analysing relativistic effects \cite{T19a}. 

Floquet theory is focusing on time-periodic systems of ordinary differential equations \cite{McL:eps,Grimshaw,Hill,Math}. It applies to space-periodic quantum systems as well, and is an alternative to Bloch's theory of space-periodic material structures. Behaviors of quantum particles in a locally periodic potential are related to behaviors of typical (time-periodic) Floquet solutions: appearances of continuous intervals of forbidden energies defining energy gaps (exponentially growing/declining solutions) and of continuous intervals of allowed energies defining energy bands (periodic/quasi period solutions). The so-called Hills equation and its implied dynamics in one dimension are classical topics in textbooks of mathematics and physics \cite{Hill}-\cite{Math}. 

As soon as the periodic potential is finite in space, a strict band/gap structure loses its meaning. Energy gap zones are no longer forbidden. Still, oscillatory transmission behaviors as function of energy seem to be localized to 'transmission bands' \cite{Dharani16}. Such bands are in this study {\it confirmed} being closely related to Floquet/Bloch energy band zones as the number of period cells is large. 
A particular feature of a transmission band is the possibility of total transmission at low energies. The number of such energies within a transmission band is related to the number of period cells of the locally periodic potential. The exact number is not clearly discussed in the reference list,  except for a resent study in \cite{T20b}.. In the present study it turns out that the number of energies causing total transmission may be different in different transmission bands. The presence of exterior potentials is found to introduce further shifts in transmission properties. 

The present study confirms that transmission bands appear separated by gap zones, where gap zone transmissions become more and more suppressed as the number of cells increases. Band/gap structures appear more pronounced at low scattering energiest. Low-energy transmission bands also show the strongest energy oscillations of transmissions. 
It is also confirmed that multi-well potentials may cause an incomplete transmission band near the threshold energy \cite{Sprung93,Dharani16}. 

The potential models of the present study have parameters chosen to avoid incomplete transmission bands. Bound states are not considered. It allows a study of quite general multi-well/barrier systems perturbed by monotonic exterior potentials.
Computations and analyses are based on amplitude-phase separations of Schr\"{o}dinger waves, originated almost a century ago \cite{Milne}-\cite{Pinney}. An unexplored flexibility of this approach is related to the many equivalent ways an amplitude function can be defined, still describing a given Schr\"{o}dinger wave function in an exact way. Applications presented make use of recent developments suitable for one-dimensional scattering \cite{T05a}, as well as aspects relevant to Floquet-type problems \cite{T20b,T19b}. Resulting formulas and conditions are exact, limited only by the accuracy of the numerical integration algorithms used.

Symmetries of the potential can be explored when defining amplitude functions in an amplitude-phase approach. For example, periodic potentials allow the existence of periodic amplitude functions, which are closely related to the Floquet/Bloch base solutions mentioned above \cite{Grimshaw}. 
Apart from computational qualities, this approach allows an analysis of transmission phenomena in terms of an intrinsic Floquet/Bloch wave amplitude and an intrinsic wave phase. These quantities contain the essential information extracted from any locally periodic potential and are independent of exact methods used. 
 
\begin{figure}
\begin{center}
\includegraphics[width=100mm,clip]{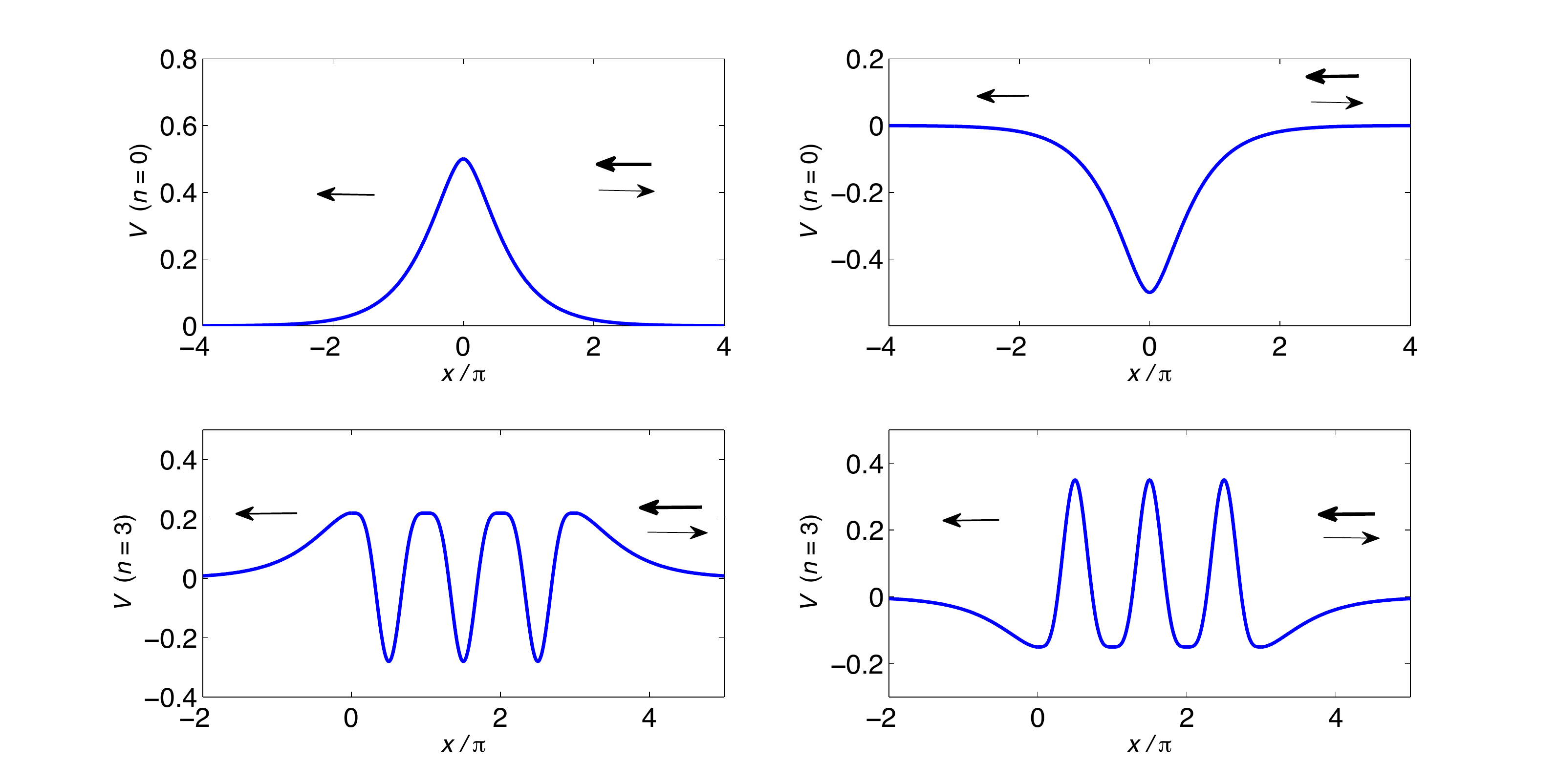}
\end{center}
\caption[ ]{\label{fig1} \small  Illustration of potentials given by (\ref{poti}),  (\ref{pote1}) and  (\ref{pote2}). For $n=0$ in the two upper subplots the potential is either a single barrier ($D=-0.5$) or a single well ($D=0.5$) type. The  lower subplots illustrate two cases where where the locally periodic potential with period length $\pi$ has three cells (denoted $n=3$). Lower left subplot corresponds to $(D,V_0)=(-0.22,-0.5)$. Lower right subplot corresponds to $(D,V_0)=(0.15,0.5)$. Arrows indicate directions of propagating wave components.}
\end{figure}

An amplitude-phase approach fits well to the Floquet/Bloch theory within energy band zones. For such energies, solutions describe oscillating quantal waves of finite norm. Independent Floquet solutions are typically written as $F(x)= P(x)\exp (\rmi\kappa x)$ and $F^*(x)= P^*(x)\exp (-\rmi\kappa x)$ \cite{McL:eps,Grimshaw}. Here  $x$ is the space coordinate, $P(x)$ is a complex, periodic function of $x$, and $\kappa$ is an energy-dependent positive constant within a band region. 
Exponential amplitude-phase type base solutions are $A_p(x) \exp [\pm\rmi\int_0^x A_p^{-2}(x')\rmd x']$, where  $A_p(x)$ is a positive, periodic amplitude function of $x$. The wave phase is defined by the amplitude function. 

The relation to Floquet/Bloch solutions can be realized by noting: that $A_p(x)=A_p(x+a)$, where $a$ is a period length, and that $\int_0^{x+a} A_p^{-2}(x')\rmd x'=\int_0^x A_p^{-2}(x')\rmd x'+\alpha$, where $\alpha=\int_x^{x+a} A_p^{-2}(x')\rmd x'$ is an $x$-independent phase. Hence, the amplitude-phase solutions satisfy  $\Psi^{(\pm)}(x+a)=\Psi^{(\pm)}(x) \exp (\pm\rmi \alpha)$. Consequently, the relations between the two wave descriptions within bands are: $\kappa =\alpha/a$ and $P(x) = A_p(x) \exp [\rmi\int_0^x \left(A_p^{-2}(x')-\alpha/a\right)\rmd x']$. Here $P(x)$ is a {\it complex}, periodic function of $x$, while $A_p(x)$ is a {\it positive}, periodic function. 

A primary effect of introducing exterior potentials is that the locally periodic part of the potential becomes shifted on the energy scale; it may lie embedded in a surrounding well or in a surrounding repulsive region; see Figure 1. This explains an additional energy shift not caused by shape parameters of wells and barriers in the locally periodic potential. Another aspect is that monotonic exterior potentials introduce additional barriers or wells, on either side of the locally periodic part of the potential.
The exterior potentials are here monotonic (repulsive or attractive) and vanish as $|x|\to\infty$. They are fitted to the locally periodic potential with continuous first-order derivatives. The potentials in Figure 1 can be characterized in two ways: (left subplots) a multi-well potential of identical cells with a symmetric repulsive exterior potential, and,  (right subplots) a multi-barrier potential of identical cells with a symmetric attractive exterior potential.

The present approach replaces the so-called 'transfer matrices' \cite{Sprung93} by so-called 'connection matrices', a tool for re-expressing different sets of linear wave functions and corresponding derivatives in terms of each other. Connections between locally valid approximate solutions often occur in semiclassical (WKB) analyses \cite{OD,JNLC84,SC}. Here, similar connections are used between globally exact fundamental solution expressions. Methods using transfer matrices and connection matrices are equal mathematical tools. Phenomena presented in this study and so-called intrinsic quantities are independent of any particular exact methods used for computing them.

Section II formulates the time-independent Schr\"{o}dinger problem of a specified total energy. The relevant wave function satisfies certain scattering conditions on an $x$-axis, from which transmission and reflection coefficients are defined. Section III presents two of many amplitude-phase approaches. It defines relevant amplitude functions and phase functions and relations between the corresponding wave representations. Relations to transmission/reflection coefficients are also derived.
Intrinsic Floquet/Bloch quantities are introduced as a third approach in Section IV. Specific formulas for analyzing transmission and reflection coefficients are derived. Quantum numbers for energy states of total transmission are suggested. Illustrations and numerical results are discussed. Band types and the fusion phenomenon are discussed and illustrated in Section V. Section VI contains concluding remarks.

\section{The quantum transmission problem in one dimension}
The time-independent Schr\"{o}dinger equation with a dimensionless space coordinate $x$ and dimensionless parameters is given by \cite{T19b}
\begin{equation}
F'' + 2\left(E-V(x) \right) F =0,
\label{ode2} 
\end{equation}
where a prime ($'$) means differentiation with respect to $x$. Equation (\ref{ode2}) is expressed as if being in atomic units. $V(x)$ represents a potential energy function that {vanishes} as $|x|\to\infty$.   $E (>0)$ represents the total scattering energy.  Equation (\ref{ode2}), with $V(x)$ being a periodic function of $x$, is a special case of a so-called 'Hill equation' \cite{Hill} . 

A truncated periodic potential in the interval $0\leq x\leq n\pi$ is fitted to attractive/repulsive exterior potentials. $\pi$ is the unit length of a period cell, and $n$ is the number of such cells.  The exterior potentials vanish as $|x|\to \infty$. The locally periodic potential is
\begin{equation}
V(x)= V_0 \sin^4(x) - D,\;\; 0\leq x\leq n\pi, \label{poti}
\end{equation}
where $D$ represents the extreme well or barrier energies caused by the two exterior potentials. Since $V(0)=V(n\pi)=-D$, the locally periodic potential appears in an attractive surrounding region for $D>0$, and in a repulsive surrounding region for $D<0$.
Symmetric exterior potential tails are introduced by
\begin{equation}
V(x)= D\left[\exp(4x)-2\exp(2x)\right], \;\; x<0, \label{pote1}
\end{equation}
 \begin{equation}
V(x)= D\left[\exp(-4(x-n\pi))-2\exp(2(x-n\pi))\right], \;\; x>n\pi. \label{pote2}
\end{equation}
The potential function and its first derivative are continuous where (\ref{poti}), (\ref{pote1}) and (\ref{pote2}) are connected.  

Two particular sets of parameter values are chosen for illustrations and numerical computations, as given in the caption of Figure 1: The parameters $(D,V_0)=(-0.22,-0.5)$ show multiple wells embedded by repulsive surrounding exterior potentials, and  $(D,V_0)=(0.15,0.5)$ show multiple barriers embedded by attractive surrounding exterior potentials. 
The two sets of parameters are chosen to avoid bound states and incomplete transmission bands. The first complete transmission bands lie within an energy range $0<E<1$. 

It is seen in Figure 1 that the number of wells and barriers are different. For $D<0$ the number of barriers is one more than the number of well. For $D>0$ the number of wells is one more than the number of barriers.

The asymptotic wave number is 
\begin{equation}
k=\sqrt{2E}, \label{kdef}
\end{equation}
and the scattering boundary conditions for a wave entering from $+\infty$ are written as
\begin{eqnarray}
F \sim t \frac{1}{\sqrt{k}}\exp (-\rmi kx), && \;\; x \rightarrow - \infty,\,\,\, \label{leftB}\\
F \sim \frac{1}{\sqrt{k}}  \exp (-\rmi kx)&&
+ \,\,\, r \frac{1}{\sqrt{k}} \exp (\rmi kx), \;\; x \rightarrow + \infty,
\label{rightB}
\end{eqnarray}
where $t$ and $r$ are the transmission and reflection amplitudes, respectively.
Transmission and reflection coefficients for symmetric potentials are defined by
\begin{equation}
T=|t|^2,\;\;R=|r|^2.  \label{RTdef}
\end{equation}

\section{General amplitude-phase approaches}
This section presents several ways to compute transmission/reflection coefficients with the aid of amplitude- and phase functions.  In particular approaches, only a single cell of the periodic part of the potential is needed for obtaining the transmission coefficient. In other approaches integration ranges across several cells, half or all the number of cells.  Contributions from the exterior parts of a symmetric potential always need particular integrations. Approaches differ by the number of amplitude functions used, and how boundary conditions for these amplitude functions are specified. All amplitude functions satisfy the same non-linear differential equation, the Milne-Pinney equation \cite{Milne}-\cite{Pinney}. For reference to intrinsic Floquet/Bloch quantities, particular periodic amplitude functions and corresponding phase functions are introduced in Section IV. Such 'intrinsic' quantities are used for analyzing detailed behaviors of transmissions.

 Amplitude functions resulting in numerically computer-time efficient formulas are introduced for each characteristic region of the potential, which requires connections of the corresponding linear waves functions. An amplitude function and its related phase function define two independent linear wave functions of the Schr\"{o}dinger problem. An arbitrary additive constant of a phase function is specified by a reference point on the $x$-axis. At a phase reference point the phase is zero. Oscillating behaviors of the linear wave function are described by the phase function. An amplitude function is a positive function of the coordinate in classically allowed potential regions. 
 
The potential region is seen as consisting of three parts; the locally periodic part, and the two exterior parts. Each of these regions is represented by a suitable amplitude-phase representation of a linear {\it fundamental solution matrix}. This $2\times 2$-matrix has a first row of two linear, independent solutions. A second row consists of the corresponding first derivatives of the first row.

Any two independent solutions of (\ref{ode2}) are defined in terms of a positive amplitude function $A(x)$ and a related real phase function $p(x)$ as \cite{T19b} 
\begin{eqnarray}
\Psi^{(\pm)}(x)=A(x) \exp (\pm\, \rmi \, p(x)),\label{ansatz}\\ p'(x) = A^{-2}(x) \;(>0), \label{Mephase}
\end{eqnarray}
where $'=\rmd/\rmd x$.
Due to the relation (\ref{Mephase}), the Wronskian determinant of the two solutions (\ref{ansatz}) is independent of $x$ \cite{T05a}. 
Any amplitude function satisfies a nonlinear Milne-Pinney equation \cite{Milne}-\cite{Pinney}
\begin{equation}
A''(x)+2\left[E-V(x) \right]  A(x)={A}^{-3}(x).
\label{Me} 
\end{equation}
Amplitude functions differ by their boundary conditions \cite{T18b}. An amplitude function is known to be more or less oscillatory due to different choices of its boundary conditions. A constant amplitude function exists in each region where the potential is constant. Several amplitude functions may be used to represent a given linear wave function. All amplitude functions formally result in exact pairs of independent solutions of the Schr\"{o}dinger equation. Different representations of a linear wave function can be expressed in terms of the others by linear combinations.

Equation (\ref{Me}) is re-written for computational purposes as a first-order differential equation as
\begin{equation}
\left[\begin{array}{c}   A(x)  \\     A'(x) \\   p(x)  \end{array} \right]' 
= \left[ \begin{array}{c}    A'(x)  \\  {A}^{-3}(x)- 2(E-V(x)) A(x) \\  A^{-2}(x) \end{array} \right]. \label{num}
\end{equation}
The integration starts at a boundary point with boundary conditions for the amplitude function. The phase function needs to be adjusted with a specified integration constant after each completed integration.

\subsection{First approach}
A fundamental solution matrix consists of $\Psi^{(\pm)}(x)$ in the upper row and $\Psi'^{(\pm)}(x)$in the lower row.  In the two exterior regions one has
\begin{equation}
{\bf \Psi}_{L,R}(x) =\left( \begin{array}{cc} A_{L,R}(x)\rme^{\rmi p_{L,R}(x)} & A_{L,R}(x)\rme^{-\rmi p_{L,R}(x)}\\
\left[A'_{L,R}(x)+\rmi A_{L,R}^{-1}(x)\right] \rme^{\rmi p_{L,R}(x)} & \left[A'_{L,R}(x)-\rmi A_{L,R}^{-1}(x)\right] \rme^{-\rmi p_{L,R}(x)} \end{array} \right), \label{extPsi}
\end{equation}
where $\det {\bf \Psi}_{L,R}(x)=-2\rmi$.  Boundary conditions for the exterior amplitudes are given by (see Ref. \cite{T05a})
\begin{equation}
A_{L,R}(x) \to k^{-1/2}, \;\; A'_{L,R}(x) \to 0, \;\; |x|\to\infty, \label{bcondL}
\end{equation}
in order to represent the well-defined amplitudes of the propagating asymptotic waves in equations (\ref{leftB}) and (\ref{rightB}). The phase reference points, where the phases vanish, are chosen at a symmetry point of the potential.
By matching these two fundamental solutions one can obtain the reflection/transmission coefficients. This approach
has a disadvantage compared to other approaches suggested. It contains less quantities for interpretations and computation times become significally larger as $n>>100$.     

To find useful espressions for the transmission coefficient one has to relate propagating wave components on either side of the locally periodic potential region. A connection between any two fundamental solutions of the Schr\"{o}dinger equation is formulated by a matrix equation involving a constant ($x$-independent) matrix. For example, the two fundamental 'exterior' solutions ${\bf \Psi}_{L,R}(x)$ in (\ref{extPsi}), represented by different amplitude functions, are related by
\begin{equation}
{\bf \Psi}_{L}(x) = {\bf \Psi}_{R}(x) {\bf \Omega}, \label{genLR}
\end{equation}
where ${\bf \Omega}$ is an $x$-independent matrix.  In the first approach, using only the exterior representations ${\bf \Psi}_{L,R}(x)$, ${\bf \Omega}$ can be determined at any matching point, say $x=x_m=n\pi/2$.  At the matching point the complex values of elements in the exterior representations (\ref{extPsi}) can be symbolically expressed as
\begin{equation}
{\bf \Psi}_{L}(x_m)=\left( \begin{array}{cc} f & f^*\\
f' & f'^* \end{array} \right),\;{\bf \Psi}_{R}(x_m)=\left( \begin{array}{cc} g & g^*\\
g' & g'^* \end{array} \right) \label{genfg}
\end{equation}
implying
\begin{equation}
{\bf \Omega}=   {\bf \Psi}^{-1}_{R}(x_m) {\bf \Psi}_{L}(x_m)=\left( \begin{array}{cc} \frac{\rmi}{2}(fg'^*-f'g^*)&\frac{\rmi}{2}(f^*g'^*-f'^*g^*) \\
\frac{\rmi}{2}(gf'-g'f) & \frac{\rmi}{2}(gf'^*-g'f^*) \end{array} \right). \label{oneO}
\end{equation}
Elements of (\ref{oneO}) can also be expressed in terms of so-called Milne-Lewis invariants, as discussed in \cite{T05a} and in \cite{T02}-\cite{Haas}. 
Symmetry properties of the  ${\bf \Omega}$-matrix in (\ref{oneO}) can be summarized as
\begin{equation}
{\bf \Omega} =  \left( \begin{array}{cc} \Delta^* & \Lambda \\
 \Lambda^* & \Delta \end{array} \right),\;\;|\Delta|^2=1+|\Lambda|^2. \label{genOmega}
\end{equation}
Numerical computations of ${\bf \Omega}$ for symmetric potentials use the phase reference point $x_m=n\pi/2$
by noting $p_L(n\pi/2)=p_R(n\pi/2)=0$. The amplitude functions $A_{L,R} (x)$ assume after integration values $A_{L,R} (n\pi/2)=w$ and $A'_{L} (n\pi/2)=-A'_{R} (n\pi/2)=w'$. Consequently, the $f, f^*$ and $g, g^*$ symbols in (\ref{genfg}) can be replaced by
\begin{equation}
{\bf \Psi}_{L}(n\pi/2) = \left( \begin{array}{cc} w &  w \\
 w'+\rmi w^{-1} &  w'-\rmi w^{-1} \end{array} \right),
\end{equation}
respectively
\begin{equation}
 \;\;{\Psi}_R(n\pi/2) =\left(\begin{array}{cc} w & w\\   -w'+ \rmi w^{-1} & -w'- \rmi w^{-1} \end{array}\right).
\end{equation}
Matrix elements $\Lambda$ and $\Delta$ of ${\bf \Omega}$ become
\begin{equation}
\Lambda = \frac{\rmi}{2}(f^*g'^*-f'^*g^*) =\frac{\rmi}{2}\left[w( -w'- \rmi w^{-1})-( w'- \rmi w^{-1})w \right]= - \rmi w w', \label{gLambda}
\end{equation}
respectively
\begin{equation}
\Delta =  \frac{\rmi}{2}(gf'^*-g'f^*)  =\frac{\rmi}{2}\left[w( w'- \rmi w^{-1})-( -w'+ \rmi w^{-1})w \right]=  1 + \rmi w w' .   \label{gDelta}
\end{equation}
Formulas are exact, but numerical errors are expected to increase as the number of cells increases. Integration extends accross the entire range of the potential. For single-barrier potentials the formulas are 'extremely' accurate; see \cite{T05a}. For a potential barrier in the cell and $D=0$, as in \cite{T05a}, $w$ and $w'$ are always positive. For a weak single well in the cell and $D=0$, $w$ is positive and $w'$ is negative. 
Some understanding of the energy dependences of $w$ and its derivative in the exterior regions is provided by the Wenzel-Kramers-Brillouin (WKB) approximation of the amplitude function, i.e. $w \approx \left[2(E-V(x))\right]^{-1/4}$, of equations (\ref{ode2}) and (\ref{Me}).

With known exact elements in ${\bf \Omega}$, scattering boundary conditions (\ref{leftB}) and (\ref{rightB}) can be explored further. In the left asymptotic region of $x$ the amplitude-phase  solution $\Psi^{(-)}_{L}(x)$ (or $f^*$ in (\ref{genfg}))  corresponds to $F(x)$ in (\ref{leftB}) as $x \rightarrow -\infty$. $\Psi^{(-)}_{L}(x)$ behaves as \cite{T05a}
\begin{equation}
\Psi^{(-)}_{L}(x) \sim  \frac{1}{\sqrt{k}}\; {\rm e}^{\rmi \delta_L}\; {\rm e}^{-\rmi kx},
\;\;\mbox{as} \;\;x \rightarrow -\infty.
\label{psiL}
\end{equation}
where $ \delta_L$ is an unspecified $x$-independent real phase.
$\Psi^{(-)}_{L}(x)$ corresponds, via (\ref{genLR}), to an equivalent expression in terms of $\Psi^{(\pm)}_{R}(x)$, given by
\begin{equation}
\Psi^{(-)}_{L}(x)=\Lambda\Psi^{(+)}_{R}(x) + \Delta \Psi^{(-)}_{R}(x),
\label{psimatch}
\end{equation}
where 
\begin{equation}
\Psi^{\pm}_{R}(x) \sim  \frac{1}{\sqrt{k}} \; {\rm e}^{\mp \rmi \delta_{R}}\;{\rm e}^{\pm\rmi kx},
 \;\;x \rightarrow +\infty.
\label{psiR}
\end{equation}
In (\ref{psiR}), $\delta_R$ is an $x$-independent real phase.
From (\ref{psimatch}) and  (\ref{psiR}) follows
\begin{equation}
\Psi^{(-)}_{L}(x) \sim   \frac{\Lambda}{\sqrt{k}} {\rm e}^{-\rmi \delta_R} {\rm e}^{\rmi kx}   
+ \,\,\,  
\frac{\Delta}{\sqrt{k}} {\rm e}^{\rmi \delta_{R}}{\rm e}^{- \rmi kx}, \,\,\, x \rightarrow + \infty.
 \label{psiRR}
\end{equation}
Note that a resonance state condition (only outgoing wave components) is $\Delta=0$. This condition is not explored in this study.
\begin{figure}
\begin{center}
\includegraphics[width=100mm,clip]{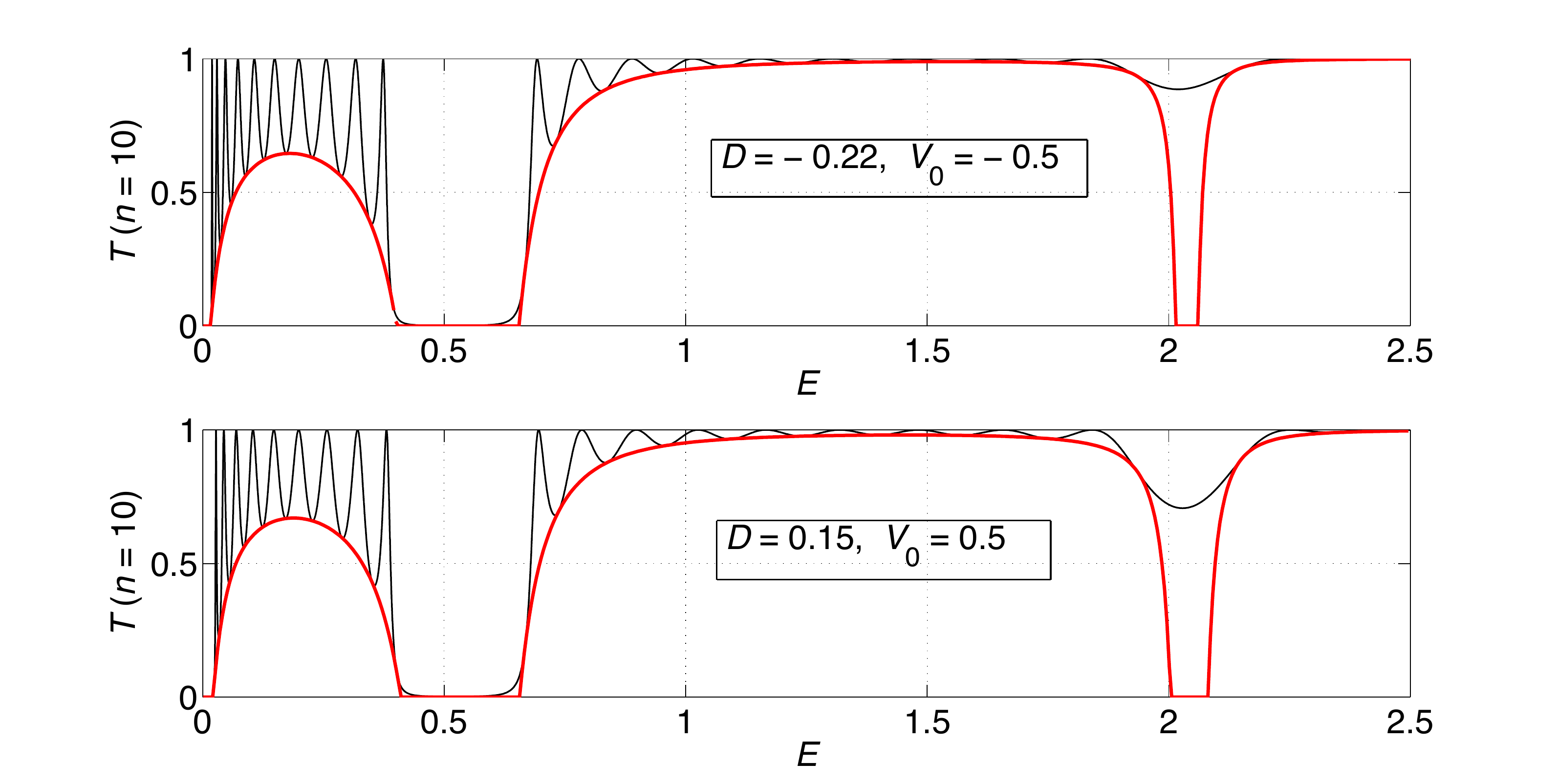}
\end{center}
\caption[]{\label{fig2} \small Illustration of transmission coefficients (thin curves) as functions of energy for potentials corresponding to the lower subplots in Figure 1 with $n=10$. Broad and narrow structures are seen. The narrow structures are located to energy ranges of broad maxima. Curves of minimal transmissions are shown as thick red curves; see formula (\ref{Tmin}).}
\end{figure}

Normalizing  (\ref{psiRR}) to agree with condition (\ref{rightB}),
the transmission and  reflection  amplitudes appear as 
\begin{equation}
t = \rme^{ \rmi (\delta_L-\delta_R)}\frac{1}{\Delta},\;\;r = {\rm e}^{-2\rmi \delta_R}\frac{\Lambda}{\Delta}.
\end{equation}
The  transmission and reflection 
coefficients defined in (\ref{RTdef}) can be expressed in terms of $\Lambda$ as
\begin{eqnarray}
T =  {\frac{1}{1+|\Lambda|^{2}}},\;\;R=  {\frac{|\Lambda|^{2}}{1+|\Lambda|^{2}}}.
\label{Tdeff1}
\end{eqnarray}
Figure 2 illustrates transmission coefficients for the potentials corresponding to the lower subplots in Figure 1, but with the number of cells $n=10$. Without exterior potentials the number of energies causing total transmission are expected to be the same in each group of peaks (contained in the transmission bands). The case  $(D,V_0)=(0.15, 0.5)$ in Figure 2 confirms the expectation of equal numbers of peaks. Hardly seen in Figure 2, peaks of total transmission for $(D,V_0)=(-0.22,-0.5)$ are different in the two transmission bands. Such a difference is not previously discussed by authors in the reference list, except in \cite{T20b} where exterior potentials are assumed being zero. 

\subsection{Second approach}
In approaches where the locally periodic part of the potential is treated separately, two phase reference points and two matching points are needed. The exterior solutions are still used outside of the periodic region.
In the region of the locally periodic potential it is preferred here to use a real-valued fundamental solution matrix composed by the real and imaginary parts of (\ref{Mephase}), i.e. by the real solutions
\begin{equation}
S(x)=A(x)\sin p(x),\;\;C(x)=A(x)\cos p(x). \label{apSCsol}
\end{equation}
$A(x)$ is now defined at the first cell boundary point $x=0$, by its boundary conditions
\begin{equation}
A_0 = 1, \;\; A'_0 =0. \label{bcondM}
\end{equation}
This choice simplifies final formulas. The fundamental solution becomes a unit matrix at $x=0$ if the phase reference point is  $x=0$, i.e. p(0) = 0.

Hence,
\begin{equation}
{\bf \Psi}(x) = \left(\begin{array}{cc}
   C(x) & S(x) \\ 
   C'(x)& S'(x)\\ 
 \end{array}\right),\;\; \det {\bf \Psi}(x) =  1,  \label{PsiM}
\end{equation}
satisfying
\begin{equation}
{\bf \Psi}(0) = \left(\begin{array}{cc}
   1 & 0 \\ 
   0 & 1
  \end{array}\right).  \label{PsiM0}
\end{equation} 
Numerical accuracy can be improved for large values of $n$ by taking into account the fundamental solution in (\ref{PsiM}).  Integration can then be restricted to one potential cell only \cite{Sprung93}, except for the integration range required for the exterior potential.  Such an approach involves a {\it principal} fundamental solution and the particular {\it monodromy} matrix:
\begin{equation}
{\bf M}(x) = {\bf \Psi}(x){\bf \Psi}^{-1}(0)= {\bf \Psi}(x),\;\;(\mbox{prinipal matrix}),\;\; {\bf M} = {\bf \Psi}(\pi){\bf \Psi}^{-1}(0)={\bf \Psi}(\pi)\;\;(\mbox{monodromy matrix}), \label{Monodef}
\end{equation}
where ${\bf M}={\bf M}(\pi)$ is the simplified notation used subsequently.  A principal solution matrix satisfies ${\bf M}(n\pi)={\bf M}^n$ \cite{Grimshaw,Math}. 

Connections with the exterior solutions are needed at $x=0$ and at $x=n\pi$. The expression of ${\bf \Psi}_{L}(x)$ in terms of the solutions ${\bf \Psi}(x)$ is
\begin{equation}
{\bf \Psi}_{L}(x) = {\bf \Psi}(x) \left[ {\bf \Psi}^{-1}(0){\bf \Psi}_{L}(0) \right]= {\bf \Psi}(x) {\bf \Psi}_{L}(0), \label{leftmiddle}
\end{equation}
where the matching point is $x=0$. At $x=n\pi$ this means
\begin{equation}
{\bf \Psi}_{L}(n\pi) = {\bf \Psi}(n\pi) {\bf \Psi}_{L}(0) = {\bf M}^n {\bf \Psi}_{L}(0). \label{middleright}
\end{equation}

The second matching at $x=n\pi$ results in a new formal expression for the connection matrix $\Omega$ in (\ref{oneO}), based on two matchings, i.e.
\begin{equation}
{\bf \Omega} = {\bf \Psi}^{-1}_{R}(n\pi){\bf M}^n{\bf \Psi}_{L}(0). \label{OM}
\end{equation}
In (\ref{OM}) the 'external' phase of the fundamental matrix ${\bf \Psi}_{R}(x)$ is defined to be zero atat $x=n\pi$.
By analyzing ${\bf \Omega}$ in (\ref{OM}), one obtains other expresions for the elements $\Lambda$ and $\Delta$, see (\ref{genOmega}).

Exterior fundamental solution expressions in  (\ref{OM}) become
\begin{equation}
{\bf \Psi}_L(0) =\left(\begin{array}{cc} v_L & v_L\\   v'_L+ \rmi v^{-1}_L & v'_L- \rmi v^{-1}_L \end{array}\right),\;{\bf \Psi}_R(n\pi) =\left(\begin{array}{cc} v_R & v_R\\   v'_R+ \rmi v^{-1}_R & v'_R- \rmi v^{-1}_R \end{array}\right),\label{LRPsi0}
\end{equation}
where now $v_L=A_L(0)$, $v'_L=A'_L(0)$, $v_R=A_R(n\pi)$, $v'_R=A'_R(n\pi)$. Symmetric exterior potentials  imply $v_R=v_L$ and $v'_R=-v'_L$. 

The monodromy matrix ${\bf M}$ is expressed in terms of amplitude and phase values corresponding to the locally periodic potential as described next. The lower subplots in Figure 1 shows two different symmetric potential cells. In both cases $A(x)$ is integrated from the boundary conditions $A(0)=1, A'(0)=0$ (at $x=0$) up to $x=\pi$. It assumes some value at $x=\pi$, say $A(\pi)=u$. Its derivative attains some value $A'(\pi)=u'$. The phase is defined being zero at $x=0$, i.e. $p(0)=0$, and becomes $p(\pi)=\beta>0$. 
The principal fundamental solution matrix (\ref{PsiM}) satisfies
\begin{equation}
{\bf M}={\bf \Psi}(\pi)=\left(\begin{array}{cc}
    u \cos \beta &  u \sin \beta \\ 
   u' \cos \beta -u^{-1} \sin \beta & u' \sin \beta + u^{-1} \cos \beta 
  \end{array}\right). \label{Mgen}
\end{equation}
Matrix powers of ${\bf M}$ needed in (\ref{OM}) are performed numerically by standard (MatLab) methods for matrix power computations. The resulting matrix ${\bf \Omega}$ then provides the transmission/reflection coefficients from the element $\Lambda = \Omega_{12}$. This completes the second amplitude-phase approach. 

For not too large $n$ (up to some hundreds of cells) an alternative second approach involves a direct integration of ${\bf M}^n$. This yields
\begin{equation}
 {\bf M}^n ={\bf \Psi}(n\pi)= \left( \begin{array}{cc} U  \cos \eta &  U  \sin \eta \\
U' \cos \eta -U^{-1} \sin \eta& U' \sin \eta +  U^{-1}\cos \eta \end{array} \right), \label{MFloqn}
\end{equation}
with replacements $u\to U=A(n\pi)$, $u'\to U'=A(n\pi)$ and $\beta \to \eta=p(n\pi)$ in (\ref{Mgen}).

The various computations of $\Lambda$ suggested consist of integrations of the non-linear Milne-Pinney equation (\ref{Me}) in the form given in (\ref{num}). After an exterior integration from $x<<1$ with particular boundary conditions, one collects amplitude values $v$ and $v'$ at $x=0$. The corresponding phase (third component in the integrated vector in (\ref{num})) is not used. Before the matrix multiplications appearing in (\ref{OM}), a separate computation is done to collect final amplitude and phase values, $U$, $U'$ respectively $\eta$, related to the locally periodic region. No further exterior integration is needed in the symmetric potential case.

\section{Approach using intrinsic Floquet/Bloch quantities} \label{Floquet}
In a third approach intrinsic Floquet quantities are defined by particular, periodic amplitude functions $A_p(x)$. Initial conditions for such amplitudes are needed. A way of finding boundary conditions for $A_p(x)$ in band zones is suggested in reference  \cite{T19b}.  To describe $A_p(x)$ in gap regions, one has to accept complex values. 

For sufficiently high energies a WKB approximation of the amplitude function is $A_p(x)\approx \left(2(E-V(x))\right)^{-1/4}$, which is periodic in the $x$-region where $V(x)$ is periodic \cite{T19b}. An exact periodic Milne solution that agrees with the WKB expression is assumed satisfying
\begin{equation}
A_p(0) =A_p(n\pi) = u_p, \;\; A'_p(0)=A'_p(n\pi) =0,\;\; p_p(0)=0, \label{bcondMp}
\end{equation}
where the value $u_p$ is so far unspecified. The phase $p_p(x)$ is related to $A_p(x)$ as given in (\ref{Mephase}). 

The fundamental solution matrix in (\ref{PsiM}) leads with the aid of (\ref{bcondMp}) to the particular values
\begin{equation}
{\bf \Psi}_p(0)= \left( \begin{array}{cc} u_p& 0 \\
0& u_p^{-1} \end{array} \right),\;\;  {\bf \Psi}_p(\pi)= \left( \begin{array}{cc} u_p \cos \alpha &  u_p \sin \alpha  \\
- u_p^{-1} \sin \alpha &  u_p^{-1} \cos \alpha \end{array} \right), \label{FMpa}
\end{equation}
where $\alpha=p_p(\pi)$. Hence, the (unique) monodromy matrix ${\bf M}$ can be expressed as
 \begin{equation}
 {\bf M}= {\bf \Psi}_p(\pi){\bf \Psi}_p^{-1}(0)= \left( \begin{array}{cc} \cos \alpha &  u^{2}_p \sin \alpha \\
 -u^{-2}_p \sin \alpha& \cos \alpha \end{array} \right).  \label{innerg}
 \end{equation}
 A periodic amplitude function $A_p(x)$ implies a simplified calculation of the powers ${\bf M}^n$. The periodicities of $A_p(x)$ and $p'_p(x)$ imply 
 \begin{equation}
 {\bf M}^n =  \left( \begin{array}{cc} \cos n\alpha&  u^{2}_p \sin n\alpha \\
 -u^{-2}_p \sin n\alpha& \cos n\alpha\end{array} \right).  \label{innergn}
 \end{equation}
 The expression in (\ref{innergn}) allows analysis of the transmission/reflection coefficients in terms of $u^2_p$ and $\alpha$, combined with an exterior quantity $v\;\;(=v_L)$ representing the exterior potentials. A single cell of the periodic part of the potential is required.
 
${\bf M}$ has real-valued elements, as found from section III. However, $u^2_p$ and $\alpha$ assume imaginary parts in the gap zones, imaginary parts that cancel.
 \subsection{Energy band regions}
The intrinsic Floquet/Bloch quantities  $u^2_p$ and $\alpha$ can be computed by boundary conditions of $A(x)$ chosen in (\ref{bcondM}). The function $A_p(x)$ can be computed once the initial value $u_p$ is known; see reference \cite{T19b}.
Monodromy matrices (\ref{Mgen}) and (\ref{innerg}) are exactly the same by uniqueness of linear solutions satisfying (\ref{ode2}) having the same boundary conditions. Hence,
\begin{equation}
 \left(\begin{array}{cc}
     u\cos \beta &  u\sin \beta \\ 
   u'\cos \beta -u^{-1} \sin \beta  &   u'\sin \beta + u^{-1} \cos \beta
  \end{array}\right) = \left(  \begin{array}{cc}
    \cos \alpha & u_p^2\sin \alpha \\ 
 -  u_p^{-2}\sin \alpha & \cos \alpha \\ 
  \end{array}  \right).  \label{MaMp}
\end{equation}
Element by element, equation (\ref{MaMp}) yields
\begin{eqnarray}
\cos \alpha &=&  u\cos \beta,  \label{rel1}\\
u_p^2\sin \alpha &=& u\sin \beta, \label{rel2}\\
 -u_p^{-2}\sin \alpha &=&  u'\cos \beta -u^{-1} \sin \beta, \label{rel3}\\
u_p^4 &=&  \frac{u\sin \beta}{ u^{-1} \sin \beta-u'\cos \beta}= \frac{u^2\sin^2 \beta}{(1-u\cos \beta)(1+u\cos \beta)}, \label{rel4}
\end{eqnarray}
where the middle member of (\ref{rel4}) is obtained by dividing (\ref{rel2}) by (\ref{rel3}). The last member of (\ref{rel4}) is obtained by eliminating $u'$ in terms of $u$ and $\beta$ from the equality of the 11- and 22-elements of (\ref{MaMp}).
The branch of the real part of $\alpha$ is taken to be the same as that of $\beta$, which is always real. Equation (\ref{rel1}) defines $\alpha$ independently of the amplitude value $u_p$.

An amplitude value $u_p^2$ is obtained from (\ref{rel4}), yielding
\begin{equation}
u^2_p= \left(\frac{u^2 \sin \beta}{\sin \beta-u'u \cos \beta}\right)^{1/2}  . \label{amprel2}
\end{equation}

Real positive values of $\alpha$ and $u_p$  specify Floquet/Bloch energy bands. 
\begin{figure}
\begin{center}
\includegraphics[width=100mm,clip]{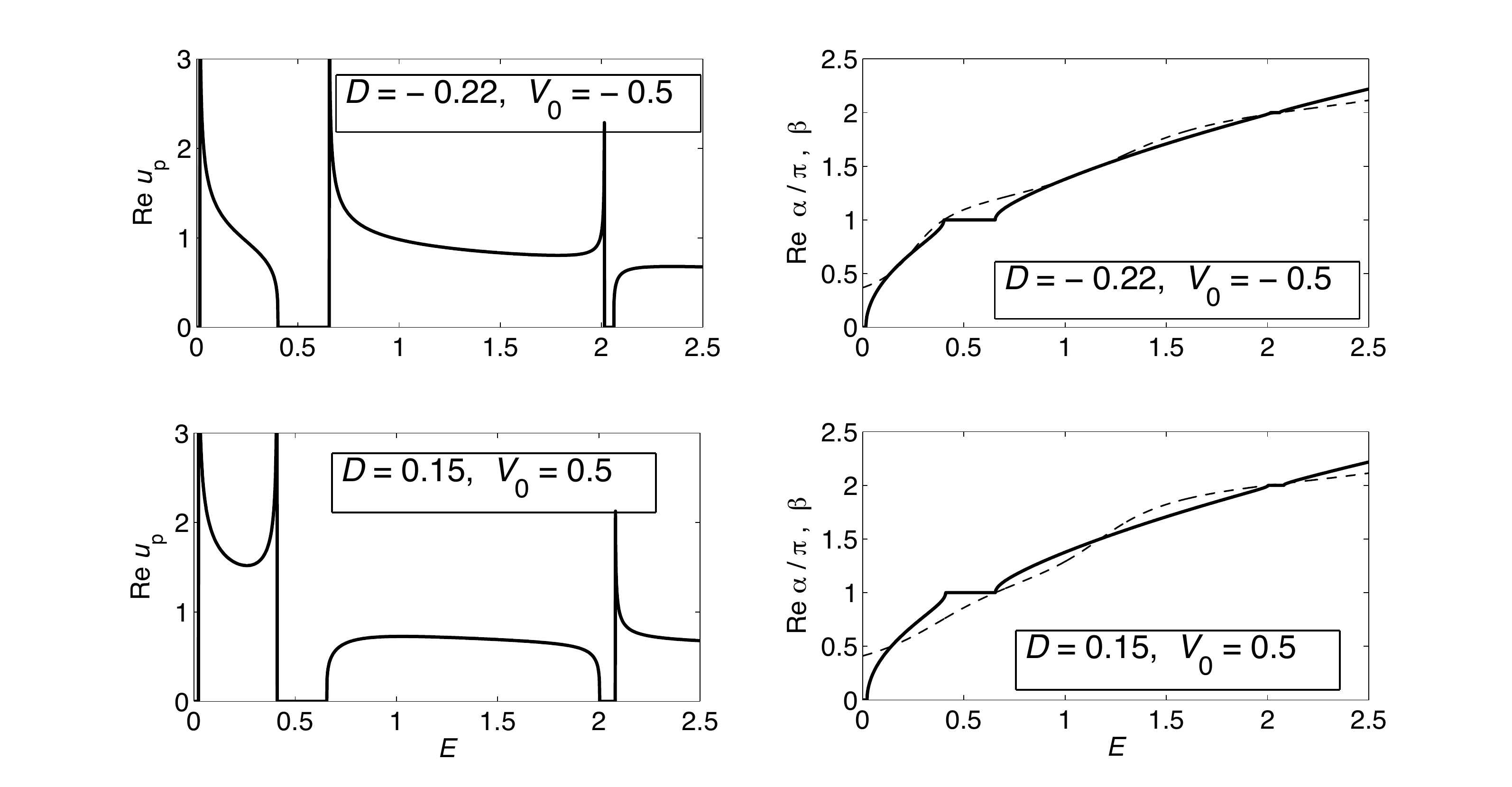}
\end{center}
\caption[]{\label{fig3} \small  Figure 3 shows the real values of the characteristic Floquet/Bloch quantities $u_p$ (left subplots), $\re \alpha/\pi$ (right subplots, solid curves) and $\beta/\pi$  (right subplots, solid dashed curves) as functions of energy. Parameters agree with those in Figure 2. The positive values of $\re u_p$ in the left subplots define Floquet energy bands. At band edges corresponding to $\re u_p=0$ an amplitude-phase solution $A(x) \sin p(x)$ is periodic, while $A(x) \cos p(x)$ is not; see reference \cite{T19b}. In the right subplots the dashed and solid curves have equal values $\alpha/\pi=\beta/\pi=1/2, 1, 3/2, \cdots$.}
\end{figure}
$u_p^2$ in the gap regions have $\im u_p^2 >0$ by choice. This implies $\im \alpha < 0$ because of the particular relation (\ref{Mephase})  between phase and amplitude. 

Figure 3 shows $\re u_p$ and $\re \alpha/\pi$ as functions of energy corresponding to the two transmission cases in Figure 2.  Transmission bands correspond to $\re u_p>0$ and gaps correspond to $\re u_p=0$. Within energy gaps $\re \alpha/\pi$ is constant.  
Within a typical energy band $\alpha$ is real. In the right subplots of Figure 3, $\re \alpha$ increases by one unit  of $\pi$ per band as function of energy. 
The figure also indicates that energy bands are associated with different behaviors of the amplitude value $u_p$ as function of energy. These different $u_p$-behaviors are important for explaining a particular phenomenon, the occurrence of different numbers of transmission peaks in different energy bands.

 A suggested way to assign a quantum number for each separate Floquet/Bloch band is
 \begin{equation}
j = {\rm Int\;} (\alpha/\pi), \;\; j=0, 1, \cdots,
\end{equation}
where $\alpha$ is real and '${\rm Int}$' means the lower integer part of a real number. It turns out that a Floquet/Bloch band may have two values of $j$ in rare situations, as discussed in subsection \ref{fusion}.

Complete band zones visible in the left subplots of Figure 3 are specified by:
\begin{eqnarray}
(D,V_0) &=& (-0.22, -0.5) \nonumber \\
\mbox{First energy band}\;\; (j=0): \;\;&& 0.0164 < E < 0.4015, \\
\mbox{Second energy band}\;\; (j=1): \;\;&& 0.6551 < E < 2.0145.
\end{eqnarray}
and
\begin{eqnarray}
(D,V_0) &=& (0.15, 0.5) \nonumber \\
\mbox{First energy band}\;\; (j=0): \;\;&&  0.0221 < E < 0.4103, \\
\mbox{Second energy band}\;\; (j=1): \;\;&& 0.6562 < E < 2.0036.
\end{eqnarray}

Band/gap edges are obtained numerically from the singularities of $u_p$, i.e. from
\begin{eqnarray}
\sin \beta = 0, \label{sin0}\;\;\;\mbox{(S)}  \label{sin1}\\
\sin \beta-u'u \cos \beta = 0. \;\;\;\mbox{(C)}\label{cos1}
\end{eqnarray}
The (S)-condition (\ref{sin1}) corresponds to a band edge where $u_p\to 0$; an energy where $A(x)\sin p(x)$ is periodic, while $A(x)\cos p(x)$ is not periodic; see \cite{T19b}. The (C)-condition (\ref{cos1}) corresponds to a band edge where $u_p\to +\infty$; an energy where $A(x)\cos p(x)$ is periodic, while   $A(x)\sin p(x)$ is not. In rare situations a gap disappears, when both conditions (\ref{sin1}) and (\ref{cos1}) are satisfied.  Then both fundamental waves $A(x)\cos p(x)$ and $A(x)\sin p(x)$ are periodic; see subsection \ref{fusion}.

Band types seen in the present study can be symbolized by (CS), (CC), (SC) and (SS), and gap types by ]CS[ and ]SC[. The order of symbols 'S/C' in this notation is: left symbol = low-energy edge type, and right symbol =high-energy edge type.  Gap zones may become small, and even vanish, even at low/moderate energies. Such behavior is illustrated in subsection \ref{fusion} below, where the potential parameter $V_0$ is varied.

 \subsection{Energy gap regions}
 The quantities $u_p^2$ and $\alpha$ have imaginary parts in the gap zones, and are computed from (\ref{rel1})-(\ref{rel4}). 
The choices of complex branches in (\ref{rel4}) respectively (\ref{rel1}) are such that $u_p^2$ gains a positive imaginary part when $u_p^4<0$, and $\alpha$ gets a negative imaginary part if $|u\cos \beta|>1$. These choices are consistent with the general relation between phase and amplitude functions in (\ref{Mephase}).  Also, the elements of the matrix (\ref{MaMp}) are real, as those in (\ref{MFloqn}) and (\ref{innergn}).
In cases the gap zones are to be analyzed in terms of $u_p^2$ and $\alpha$, one puts $u_p^2=\rmi |u_p^2|$ and $\alpha=\re \alpha - \rmi |\im \alpha|$. Then the matrix ${\bf M}^n$ in (\ref{innergn}) is well defined.

The intermediate $x$-values of the amplitude function $A_p(x)$ and of $p_p(x)$ are difficult to calculate directly from the Milne-Pinney equation (\ref{Me}).  Peculiar singular behaviors occur; e.g. real/imaginary parts of $A_p(x)$ turn out to have cycloidal form with real and imaginary parts of $A_p(x)$ approaching zero. However, the intermediate functions  $A_p(x)$ and of $p_p(x)$ are not needed. Only  the values $u_p^2$ and $\alpha$ define the matrix (\ref{innergn}) for any number of cells.

\subsection{ Analysis of ${\bf \Omega}$ and $T$ in energy bands}
The matrix ${\bf M}^n$ in (\ref{innergn}) together with  (\ref{LRPsi0})  is used to express the matrix ${\bf \Omega}$ in (\ref{OM}) in closed form. 
Formulas are simplified by the symmetry assumption of the exterior potentials by $v_R=v_L=v$ and $v'_R=-v'_L=-v'$ in (\ref{LRPsi0}). Matrix elements of ${\bf \Omega}$ are specified by new expressions for $\Lambda$ and $\Delta$, where $\Lambda=\Lambda_p$ is given by
\begin{equation}
\Lambda_p = -\rmi vv'\,\cos n\alpha+\frac{\rmi}{2} \left[\frac{v^2}{u_p^2}-\frac{u_p^2}{v^2}\left[1+(v v')^2\right] \right] \sin n\alpha.  \label{Lambda}
\end{equation}
The transmission coefficient $T$ can be expressed as
\begin{equation}
T= \frac{1}{1+\left|\Lambda_p\right|^2}. 
\end{equation}
$\Lambda_p$ agrees with the corresponding element in (\ref{gLambda}) for $n=0$, since $w=v$ being the same exterior amplitude value at the matching point $x=0$. 

With vanishing exterior potentials ($D=0$) and $n \geq 1$, the expression $\Lambda_p$ reduces to
\begin{equation}
\Lambda_{p, D=0} =\frac{\rmi}{2} \left[\frac{v^2}{u_p^2}-\frac{u_p^2}{v^2}\right] \sin n\alpha, \label{LambdaD0}
\end{equation}
since $v'=0$ with  $v=k^{-1/2}$ in this limit.
The zeros of $\Lambda_{p, D=0}$ predict peaks of total transmission within each energy band, the sharp peaks due to zeros of $\sin n\alpha$. Broad transmission maxima are due to minima of the factor $\frac{1}{2} \left[\frac{v^2}{u_p^2}-\frac{u_p^2}{v^2}\right]$, which may obscure sharp energy peaks of  $T$. The case of vanishing exterior potentials is discussed in more detail in \cite{T20b}.

The expression $\Lambda_p$ in (\ref{Lambda}) can be written
\begin{equation}
\Lambda_p = \rmi \left( J_p \sin n\alpha  - J_{X}\,\cos n\alpha\right) \label{LambdaLP}
\end{equation}
with
\begin{equation}
J_{X}= v v', \;\;J_p = \frac{1}{2} \left[\frac{v^2}{u_p^2}-\frac{u_p^2}{v^2}\left[1+(v v')^2\right] \right]. \label{Jdef}
\end{equation}
The quantities $J_{X}$ and $J_p$ are independent of the number $n$ of cells. $J_p$ is real and finite only in band regions.  $J_{X}$ and $J_p$ determine broad energy structures of transmission.  They tend to vanish for high-energy bands, implying that the high-energy transmission coefficient $T$ approaches unity. At an (S)-edge: $J_p\to+\infty$ from (\ref{Jdef}), since $u_p \to0$ from (\ref{sin1}). At a (C)-edge: $J_p\to-\infty$ from (\ref{Jdef}), since $u_p \to+\infty$ from (\ref{cos1}). These behaviors of $J_{X}$ and $J_p$ are indicated in Figures 4 and 5.

Knowing the band edges, the quantities $J_{X}$, $J_p$ and $\im \Lambda_p$ make it  possible to analyze the transmission coefficients in Figure 2. The top subplot of Figure 4 shows $\Lambda_p$ as functions of energy in the first respective transmission bands in Figure 2.  The solid line is for $(D,V_0)=(-0.22,-0.5)$  and the broken line is for $(D,V_0)=(0.15,0.5)$. The curves have 10 respectively 9 zeros. The number of cells is the same, but the numbers of barriers/wells are different. However, the key difference between the repulsive/attractive exterior models is related the $J_p$-behaviors seen in the middle subplot of Figure 4. The solid (repulsive exterior potential) $J_p$-curve has a sign change, while the dashed (attractive exterior potential) curve stays negative. The solid (repulsive exterior potential) $J_{X}$-curve in the bottom subplot has a definite, positive sign, while the dashed curve is negative.  The  $J_{X}$-behaviors are explained by their first-order WKB expressions $\left[2(E-V(0))\right]^{-1/4}$ not too close to the classical turning point where $E+D = 0$.
\begin{figure}
\begin{center}
\includegraphics[width=100mm,clip]{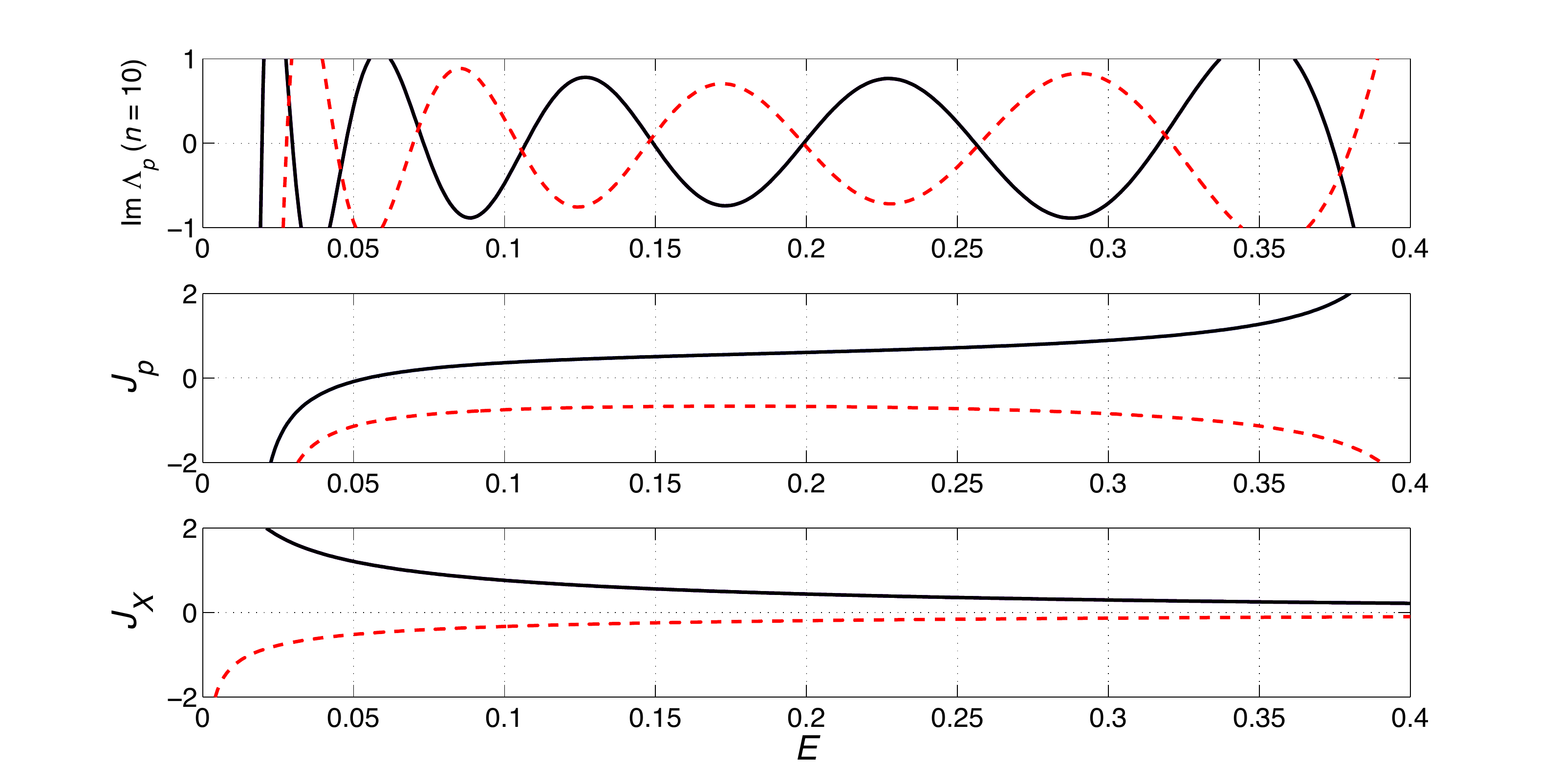}
\end{center}
\caption[]{\label{fig4} \small Analysis of the group of peaks in Figure 2 corresponding to the first band $j=0$. The top subplot shows $\Lambda_p$ in (\ref{LambdaLP}) as function of $E$. The solid black curve corresponds to $(D,V_0)=(-0.22,-0.5)$ and the red broken curve corresponds to $(D,V_0)=(0.15,0.5)$. The middle and bottom subplots show the corresponding energy behaviors of $J_p$ respectively  $J_{X}$ as functions of energy. Note that the solid  $J_p$-curve changes sign near $E=0.05$.}
\end{figure}
\begin{figure}
\begin{center}
\includegraphics[width=100mm,clip]{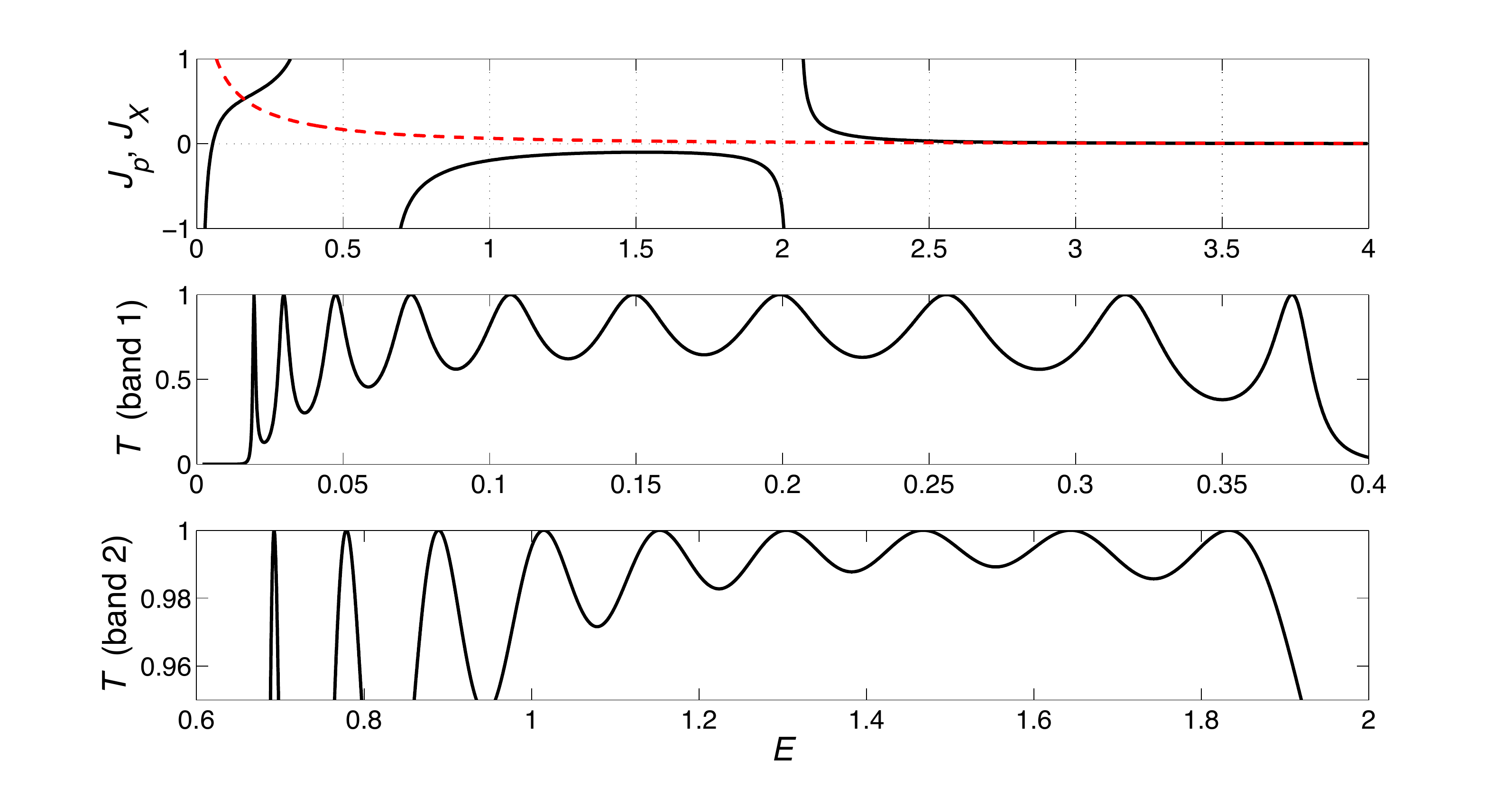}
\end{center}
\caption[]{\label{fig5} \small Analysis of the first two energy bands for the potential with parameters $(D,V_0)=(-0.22,-0.5)$. The top subplot shows $J_p$ (solid curve) and $J_{X}$ (broken curve) as functions of energy. Three bands are defined by the three visible solid curves. The third band in the top subplot shows $J_p \approx J_{X}  \approx 0$, which means that oscillations of $T$ as function of energy are too small to illustrate clearly. The transmission oscillations in the first and second bands are illustrated in the two lower subplots. The first band has 10 energies of of total transmission, while the second band has 9 energies of total transmission.}
\end{figure}

Transmission oscillations within bands $j=$0 and 1 of the repulsive model, $(D,V_0)=(-0.22,-0.5)$, are magnified in Figure 5; see also Figure 2. In the top subplot $J_p$ (solid curve) and $J_{X}$ (dashed curve) are shown in an energy range extending beyond the two first bands $j=0$ and 1. The middle subplot shows the ten peaks of total transmission of the first band. The bottom subplot shows the second energy band containing nine peaks of total transmission. This difference is analyzed in more details in subsection \ref{Peaks}.
\subsection{Minimal transmission}
$J_{X}$ and $J_p$ are key quantities in predicting minimal transmissions within energy bands.
Minimal transmissions for any possible number of cells occur when $|\Lambda_p|$ is maximal. From (\ref{LambdaLP}) one finds 
\begin{equation}
\left|\Lambda_p\right|^2_{max} = J_p^2 + J_{X}^2.\;\;\mbox{(energy band regions)} \label{Jmin}
\end{equation}
The formula is not valid at band edges and in gap regions.  $J_p$ is singular at band edges, a singularity that in a more detailed analysis is cancelled by the factor $\sin n\alpha$ in (\ref{LambdaLP}). Minimal transmissions inside energy bands are equal to an $n$-independent energy function given by
\begin{equation}
T_{min}= \frac{1}{1+\left|\Lambda_p\right|^2_{max}}. \;\;\mbox{(energy band regions)} \label{Tmin}
\end{equation}
In Figure 2, the energy gaps are represented by a minimum transmission approximated by $T_{min}=0$. This limit is only correct as $n\to \infty$. The solid, red curve in Figure 2 shows $T_{min}$ as function of energy.  

For low values of $n$  one can find peaks of total transmission also in gap regions (not illustrated); see Table I and details given in subsection \ref{genpeaks}.

\subsection{Energy conditions and quantum numbers for total transmission} \label{Peaks}
In a first energy band ($0< \alpha< \pi$) specified by a quantum number $j=0$, one has $0< n\alpha<n\pi$, which allows several oscillation of $\sin n\alpha$ and $\cos n\alpha$ in $\Lambda_p$, equation (\ref{LambdaLP})  in the band.  
To single out individual zeros of $\Lambda_p$, this interval of $n\alpha$ is divided into subintervals, each
containing one of the 'intermediate' values $n\alpha = \pi, 2\pi,\cdots , (n-1) \pi$. These values of $n\alpha$ are not singular. They do not contain the band edge values. 

The second band ($j=1$), defined by $\pi< \alpha< 2\pi$, is associated with phases $n\pi< n\alpha< 2n\pi$, and so on. Hence, a $j$th band is defined by phases in the interval $jn\pi< n\alpha< (j+1)n\pi$, $j=0, 1, \cdots$. Each band is divided into subintervals containing integer multiples of $\pi$ useful for assigning quantum numbers for a band-specific total transmission.

For large numbers of cells, the energy functions $J_{X}$ and $J_p$ become almost constants in the small energy subintervals just mentioned, while $n\alpha$ varies within a unit of $\pi$. The case of a constant exterior potential ($J_{X}=0$) implies $\tan n\alpha =0$, which is satisfied by $n\alpha=\mbox{integer}\times \pi$. The allowed integers in the first band ($j=0$ and $0< n\alpha< n\pi$) are $n\alpha=\nu\pi$, with $\nu=1, 2, \cdots, n-1$. This implies $n-1$ peaks of total transmission due to $\sin n\alpha$.

The second band ($j=1$ and $n\pi< n\alpha< 2n\pi$) contains 'peak' phase solutions $n\alpha=(n+\nu)\pi$, with $\nu=1, 2, \cdots, n-1$, and so on. In general, the $j$th band has 'peak' phase solutions $n\alpha=(jn+\nu)\pi$ with the same number of $\nu$-values. 

Zeros of  $J_p$ need to be considered in case $J_{X}=0$ \cite{T20b}.
However, an exact treatment of (\ref{LambdaLP}), with $J_{X}\neq0$,
 leads to
\begin{equation}
n\alpha - \tan^{-1}\frac{J_{X}}{J_p} = (jn+\nu)\pi,\;\;-\pi/2< \tan^{-1}\frac{J_{X}}{J_p} < \pi/2, \label{npzero}
\end{equation}
where $J_{p}\neq0$.
Condition (\ref{npzero}) expresses zeros of $\Lambda_p$ in (\ref{LambdaLP}) within band zones. It predicts all but one  entry in Table I, an energy state that does not belong to an energy band; see subsection \ref{genpeaks}. Double $\nu$-entries in Table II in the first transmission band ($j=0$) are caused by the sign change of $J_{X}/J_p$; one with a negative value of $J_{X}/J_p$, assigned ($\nu-$), and one with a positive value of $J_{X}/J_p$, assigned ($\nu+$). The $\pm$-notation is needed only for the value of $\nu$ affected by the sign change of $J_p$ and if $J_{X}\neq 0$.

A zero located numerically with the aid of the energy function $\im \Lambda_p$ from (\ref{LambdaLP}), provides quantum numbers $j$ and $\nu\pm$ according to
\begin{equation}
j={\rm Int}\, (\alpha/\pi),\;\; \nu = n(\alpha/\pi-j)-\frac{1}{\pi}\tan^{-1}\left(\frac{J_{X}}{J_p}\right),\;\; 
\pm = \sgn \left(\frac{J_{X}}{J_p}\right), \label{q-nr}
\end{equation}
where the number $n$ of cells is given.

\subsection{General computation of $\Lambda$} \label{genpeaks}
An energy of total transmission not predicted by $\Lambda_p$ is found in Table I for $n=2$. This energy is confirmed by direct calculations of $\Lambda$ and its zeros, not depending on periodic amplitude functions. Such calculations are based on the equality of ${\bf M}^n$  in (\ref{MFloqn}) and ${\bf M}^n_p$  in (\ref{innergn}). Hence, the relatively compact expression of $\Lambda_p$ can be generalized to an  expression that is independent of the periodic amplitude function.  A formally exact $\Lambda$-expression for $n \geq1$ follows from the substitutions
\begin{equation}
\cos n\alpha \to U \cos \eta \;(\mbox{or}\;U'\sin \eta+U^{-1}\cos \eta),\;\;u_p^2 \sin n\alpha \to U \sin \eta,\;\; u_p^{-2} \sin n\alpha \to U^{-1} \sin \eta - U' \cos \eta,
\end{equation}
in the $\Lambda_p$-expression (\ref{Lambda}). As a result,
\begin{equation}
\Lambda  = -\rmi vv'\,U\cos \eta+\frac{\rmi}{2} \left\{v^2 \left(U^{-1} \sin \eta - U' \cos \eta \right) -v^{-2}\left(1+(v v')^2\right)U \sin \eta \right\} \label{LambdaUP}
\end{equation}
or equivalently
\begin{equation}
\Lambda  = -\rmi \left\{vv' U^{-1} +\frac{1}{2} v^2 U'\right\} \cos \eta +\rmi  \left\{ \frac{1}{2} \left(v^2 U^{-1}-Uv^{-2}(1+(v v')^2\right)-vv' U' \right\} \sin \eta. \label{LambdaUP2}
\end{equation}
Formulas (\ref{LambdaUP}) and  (\ref{LambdaUP2}) are exact and valid at any energy. However, application of these formulas are computationally more time consuming than computations based on (\ref{Lambda}) or (\ref{LambdaLP}). Integrations of $U$ extend over the the entire periodic range $0\leq x\leq n\pi$. Unfortunately, the factors multiplying $\sin \eta$ and $\cos \eta$ depend on the number of cells and do not generalize $J_p$ and $J_{X}$ as $n$-independent analytic tools.

Energy behaviors of $\Lambda$ from (\ref{LambdaUP}) (or  (\ref{LambdaUP2})) and $\Lambda_p$ are the same within energy bands. Since (\ref{LambdaUP}) and  (\ref{LambdaUP2}) are general, one has a tool to differentiate band/gap entries in Tables I and II; those belonging to a Floquet/Bloch band, and those not belonging to a Floquet/Bloch band. Only $\Lambda$ from (\ref{LambdaUP}) (or  (\ref{LambdaUP2})) predicts the peak energy $E_{2,0,0}$ in Table I. 

\begin{table}
\begin{center}
\small
\begin{tabular}[t]{rrcc} \hline
\hline
\hline
&$(n,j,\nu)$&&$E_{n,j,\nu}$ \\
\hline
\hline
\hline
\hline
&$(2,0,0)$&&0.001970$^*$\\
&$(2,0,1)$&&0.167532\\
\hline
&$(10,0,1)$&&0.027824\\
&$(10,0,9)$&&0.380714\\
\hline
&$(10,1,1)$&&0.695647\\
&$(10,1,9)$&& 1.841595\\
\hline
&$(100,0,1)$&&0.022141\\
&$(100,0,99)$&&0.409948\\
\hline
&$(100,1,1)$&&0.656608\\
&$(100,1,99)$&&1.998824\\
\hline
\hline
\end{tabular}
\caption{\small Selected values of $E_{n,j,\nu}$ for $(D,V_0)=(0.15, 0.5)$. $^*$The entry $E_{2,0,0}$ is obtained from zeros of (\ref{LambdaUP}).}
\end{center}
\label{table1}
\end{table}
\begin{table}
\begin{center}
\small
\begin{tabular}[t]{rrcc} \hline
\hline
\hline
&$(n,j,\nu)$&&$E_{n,j,\nu}$ \\
\hline
\hline
\hline
\hline
&$(2,0,1-)$&&0.049619\\
&$(2,0,1+)$&&0.190902\\
\hline
&$(10,0,1)$&&0.019615\\
&$(10,0,3-)$&&0.047445\\
&$(10,0,3+)$&&0.073183\\
&$(10,0,9)$&&0.373860\\
\hline
&$(10,1,1)$&&0.692562\\
&$(10,1,9)$&& 1.833093\\
\hline
&$(100,0,1)$&&0.016419\\
&$(100,0,28-)$&&0.052928\\
&$(100,0,28+)$&&0.055593\\
&$(100,0,99)$&&0.401141\\
\hline
&$(100,1,1)$&&0.655576\\
&$(100,1,99)$&&2.007129\\
\hline
\hline
\end{tabular}
\caption{\small Selected values of $E_{n,j,\nu}$ for $(D,V_0)=(-0.22, -0.5)$. The double $\nu$-entries in the first transmission band ($j=0$) are caused by the sign change of $J_{X}/J_p$; see the middle subplot in Figure 4. As a result there are $n$ peaks of total transmission in the first band for $n=2,10, 100$.}
\end{center}
\label{table2}
\end{table}
\begin{figure}
\begin{center}
\includegraphics[width=100mm,clip]{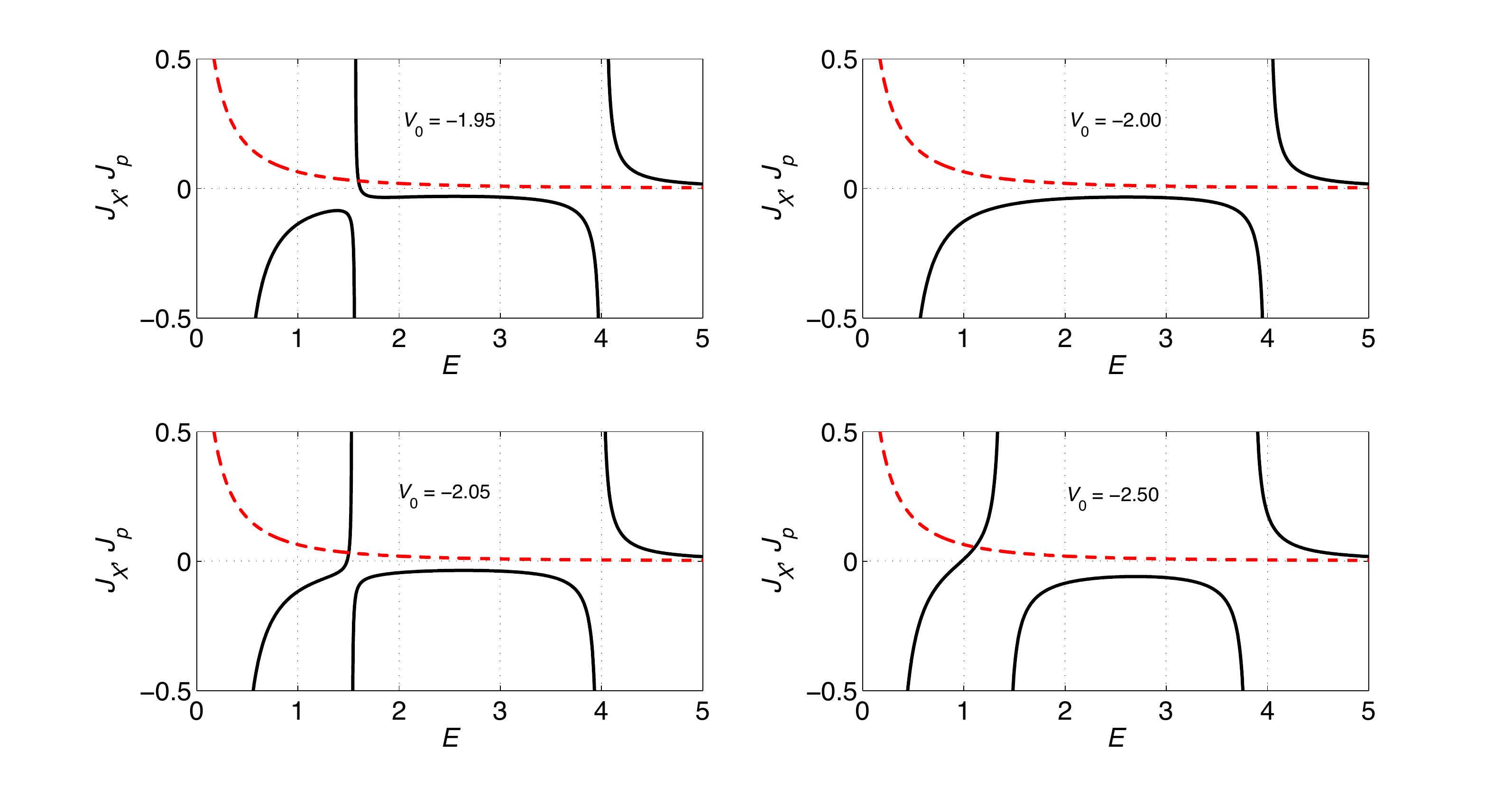}
\end{center}
\caption[]{\label{fig6} \small The subplots from top left to bottom right show $J_p$ (solid curve) and $J_{X}$ (broken curve) as functions of energy for parameters $D=-0.22$ together with respectively $V_0 =$ -1.95, -2.00, -2.05 and -2.50. The top right subplot shows a combined energy band. See also Figure 7.}
\end{figure}
\begin{figure}
\begin{center}
\includegraphics[width=90mm,clip]{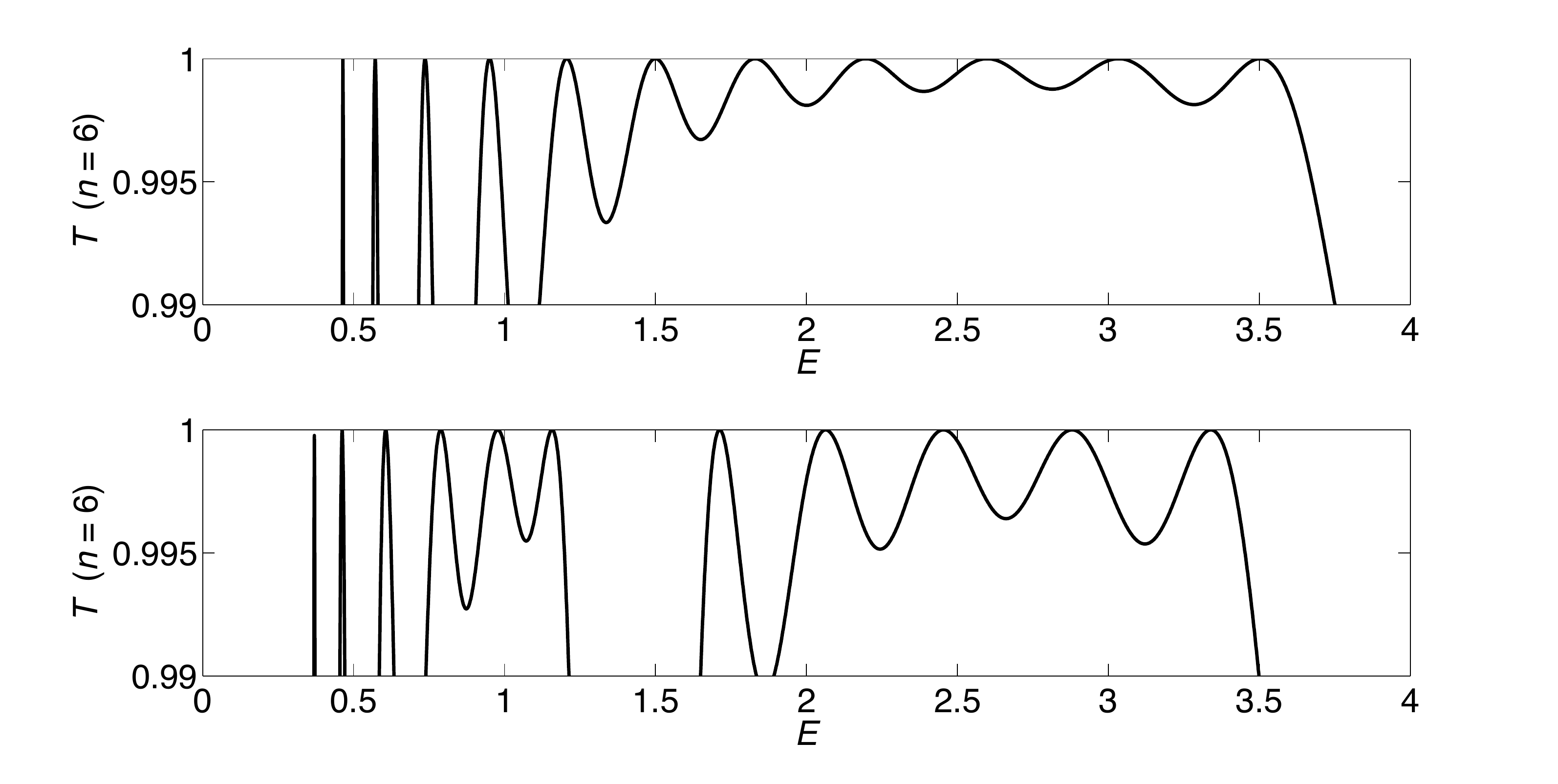}
\end{center}
\caption[]{\label{fig7} \small The figure shows $T$ as function of energy for $n=6$ and $D=-0.22$, together with $V_0 =$ -2.00 (upper subplot) respectively $V_0 =$ -2.50 (lower subplot). }
\end{figure}
\section{Band types and a band fusion phenomenon} \label{fusion}
It seems that band types and energy orders of such types  need to be calculated for each potential model. Band types are here related to $V_0$ of the periodic part of the potential and are not affected by exterior potentials. Band/gap locations on the energy scale are simply shifted due to the potential parameter $D$.

As the potential parameter $V_0$ is changed, band/gap regions of various types are modified. Gaps may dissappear and a phenomenon of 'band fusion' occur. A trivial value of $V_0$ causing 'band fusion' is $V_0=0$. All bands fuse and no gaps exist. Non-trivial fusion phenomena have not been observed for the model with attractive surrounding potentials in this study.
To illustrate a non-trivial case for the model with exterior repulsive potentials, the potential parameter $V_0$ is changed to a sequence of more negative values (deeper wells), illustrated in Figure 6. Band zones are characterized by energy curves of $J_p$ for each value of $V_0$. $D=-0.22$ is not changed. 

A {\it gap} zone near $E \approx 1.55$ of the type ]CS[ is appears for $V_0=-1.95$ in the top left subplot of Figure 6. For $V_0=-2.00$ (top right subplot), the gap has disappeared and a fused band is seen. For $V_0=-2.05$ (bottom left subplot) the gap reappears as a ]SC[-type gap. For  $V_0=-2.50$ (bottom right subplot) this gap zone has increased in size. The final (CS)-band in this subplot is similar to the ($j=0$) 'band 1' in Figure 5.  However, the first low-energy transmission band seen in Figure 6 has quantum number $j=1$. The fused band for $V_0=-2.00$ is assigned two quantum numbers by $j=(1,2)$, both $j=1$ and $j=2$ according to (\ref{q-nr}). Consequently, $j=1$ defines only an energy subset of a fused band.

Figure 7 shows the transmission coefficients as functions of energy for $n=6$. The potential parameters are $(D,V_0)=(-0.22,-2.00)$ (upper subplot) and $(D,V_0)=(-0.22,-2.50)$ (lower subplot).  These parameters correspond to the top/bottom right subplots of Figure 6.  The upper subplot in Figure 7 represents the fused band containing 11 peaks of total transmission. The lower subplot in Figure 7 shows 6 peaks in the separated (CS) band and 5 in the separated (CC) band. For $E=-1.95$ (not illustrated) the number of peaks are 5 in the (CC) band and 6 in the (SC) band. 
\begin{figure}
\begin{center}
\includegraphics[width=100mm,clip]{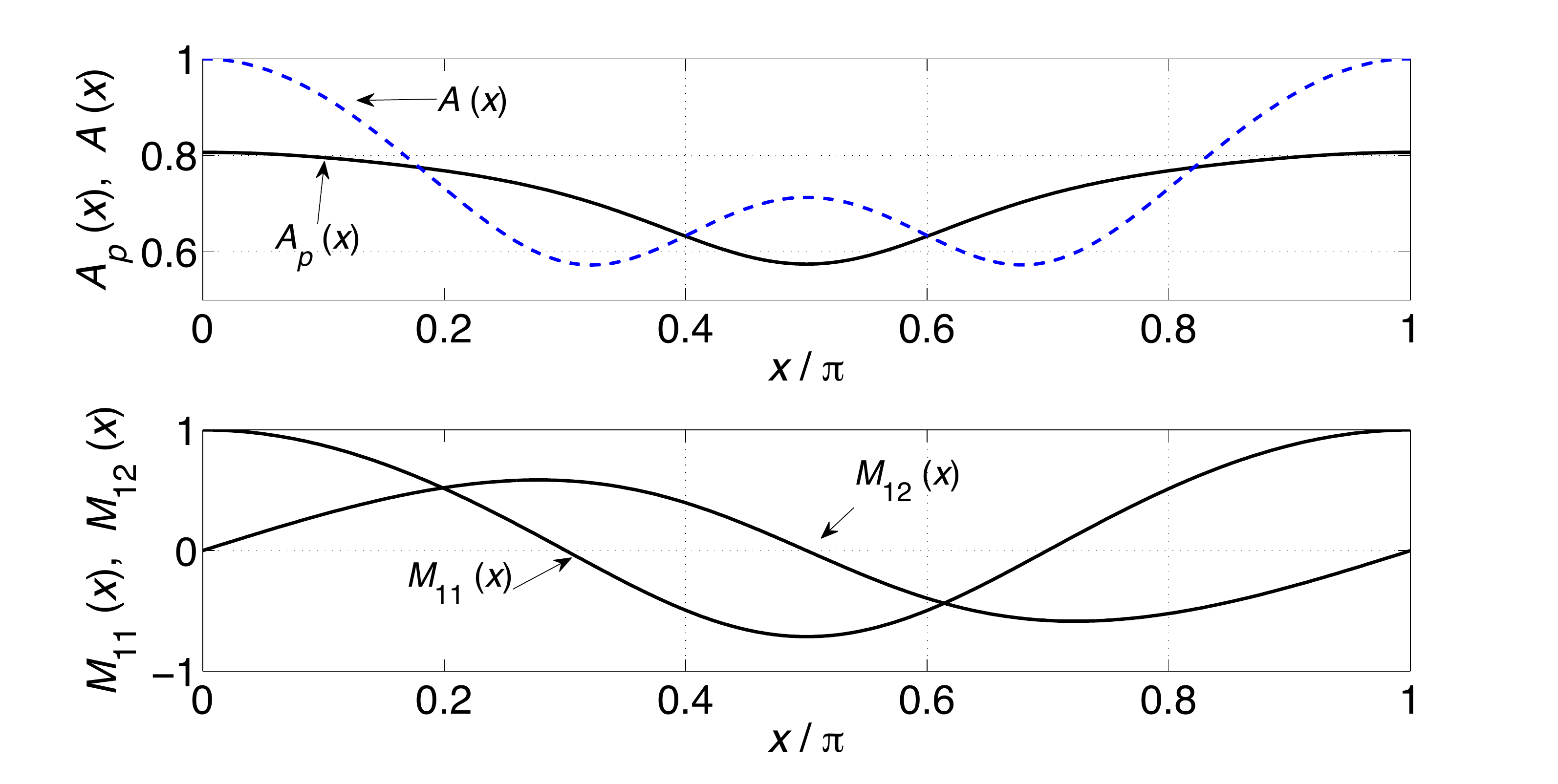}
\end{center}
\caption[]{\label{fig8} \small The figure shows two periodic amplitude functions (upper subplot) and the corresponding principal solutions (lower subplot) as a an energy gap has disappeared for $E=1.55$ and $(D,V_0)=(-0.22,-2.00)$. Although the amplitude functions are different, they represent the same linear principal solutions  in the locally periodic region.}
\end{figure}

From conditions (\ref{sin1}) and (\ref{cos1}) it follows that a fused band exists whenever the two band/gap edges disappear, i.e. for $\sin \beta = 0$ and $uu'=0$ at $E\approx1.55$; the singular behavior of $u_p$ is cancelled.  A numerical investigation shows that $u=1$ and $u'=0$. Hence, the amplitude function $A(x)$ is periodic here. However, formula (\ref{amprel2}) gives the numerical result $u_p=0.8061$ for  $A_p(x)$. The situation of two existing periodic amplitude functions is not familiar to the author and may raise doubts about the amplitude-phase representations of unique linear Schr\"{o}dinger solutions. A check of the uniqueness is illustrated in Figure 8. The principal (Schr\"{o}dinger) solutions for $E=1.55$ satisfying the same boundary conditions with two different amplitude functions, are represented by:
\begin{eqnarray}
M_{11}(x) &&= A(x) \cos p(x),  \;\;\mbox{respectively}\;\; = \left(A_p(x)/u_p\right) \cos p_p(x) \\
M_{12}(x) &&= A(x) \sin p(x),  \;\;\mbox{respectively}\;\; =u_p A_p(x)  \sin p_p(x),
\end{eqnarray}
with boundary values at $x=0$ given by
\begin{equation}
A_p(0)=u_p=0.8061456,\;\; A(0)=1,\;\; A'_p(0)=A(0)=0, \;\; p_p(0)=p(0)=0.
\end{equation}
The phase values at $x=\pi$ are
\begin{equation}
\alpha=\beta= 2\pi.
\end{equation}
 
Figure 8 illustrates the two periodic amplitude functions and their corresponding (Schr\"{o}dinger) normalized (principal) solutions as functions of $x$. The solutions turn out to be the same, whether one uses $A(x)$ or $A_p(x)$.
Although these amplitude functions are different, they represent the same pair of solutions. Furthermore, both Schr\"{o}dinger solutions are periodic. Both of them would not be periodic at a band/gap edge. At other energies within the fused band the amplitude $A(x)$ is not periodic, but $A_p(x)$ still is. 
It is not clear whether the condition $v'=0$ at $x=\pi$ strictly impies that $A(x)$ is periodic, or not. 
\begin{table}
\begin{center}
\small
\begin{tabular}[t]{rccc}
\hline
\hline
$p$&$(j, j+1)$&$V_f$&$E_f+D$ \\
\hline
\hline
4&(1,2)&-2.0000 &1.3310\\
4&(2,3)&-8.0000 &1.9804\\
\hline
6&(1,2)&-3.5599 &1.1032\\
6&(2,3)&-1.5982 &4.0211\\
\hline
8&(1,2)&-5.0114 &0.9860\\
8&(2,3)&-3.1364 &3.7140\\
\hline
\end{tabular}
\caption{\small Fused bands are denoted $(j, j+1)$, where the relevant band numbers are $j=1$ and 2. Fusion energies are expressed as $E_f+D$, and the corresponding potential parameters as $V_f$.  Entries refer to the class of potentials $V(x)=V_0\sin^{q}(x)-D$, with $q=4, 6, 8$.}
\end{center}
\label{table3}
\end{table}

Fusion phenomena are rare. Particular potential parameters are required at particular energies. For potential cells of the analytic form $V(x)=V_0\sin^{q}(x)-D$, where $q=4,6$, and 8, some specific fusion values of energy, $E=E_f$, and $V_0=V_f$ are calculated and collected in Table III. Band fusion for $q=2$ is not observed for unclear reasons. The two conditions for a fused band ($j, j+1$) are: $\beta=(j+1)\pi$ and $v'=0$.

Entries in Table III show $V_0=V_f$ associated with fusion phenomena of bands $j=(1,2)$ and $j=(2,3)$. The corresponding energies are expressed as values of $E_f+D$, since $D$ just shifts the energy scale. No positive values of $V_f$ have been found. An entry corresponding to a fused band $(j,j+1)$ is obtained by a sequence of Newton iterations. The additional condition $v'=0$ required is obtained by varying $V_0$ such that the condition $\alpha=(j+1)\pi$ is maintained. The latter condition is obtained by varying $E$ for a given $V_0$. In this double iteration process one finally finds $V_f$ and $E_f+D$ for a band fusion.

To summarize: As gap edges coalesce, the amplitude $A_p(x)$ becomes periodic instead of being singular. A possible second periodic amplitude may exist. A larger band  is formed, allowing a doubled range of the intrinsic (positive) phase $\alpha$, and consequently $n\alpha$. 

\section{Concluding remarks} 
Exact formulas for transmission/reflection coefficients are derived for potentials with a locally periodic part and symmetric exterior potentials. Various amplitude-phase approaches are suggested.
Transmission analysis in terms of a characteristic periodic amplitude and a characteristic phase is explored in two detailed case studies. These quantities are closely connected to basic Floquet/Bloch solutions. Quantum numbers for peaks of total transmission are suggested. 

Detailed numerical computations and graphical illustrations are presented for two potential models: one with a posive energy shifted multi-well and one with a negative energy shifted multi-barrier potential. Floquet/Bloch zones are found to change character for the multi-well model as the strength parameter is changed. Band/gap edges may cross, and at particular parameter values of the potential and the energy edges disappear. A gap zone then becomes a regular 'inner' point of a 'fused' band. 

Different band types are found to imply different numbers of energy peaks of total transmission. As neighboring bands combine (fuse) to a single band, respective numbers of energy peaks of total transmission add. This phenomenon is illustrated for the model a multi-well potential attached to repulsive exterior potentials. For attractive exterior potentials attached to a multi-barrier potential all transmission bands have the same number of peaks of total transmission, one less than the number of cells.


\end{document}